\documentclass[aps,pra,twocolumn,superscriptaddress]{revtex4-1}

\usepackage{array}
\usepackage{graphicx}
\newcommand{\tn}{\tabularnewline}
\newcolumntype{x}[1]{%
	>{\centering\hspace{0pt}}p{#1}}%
\newcolumntype{y}[1]{%
	>{\raggedright\hspace{0pt}}p{#1}}%

\usepackage{amsfonts}
\usepackage{amsmath}
\usepackage{amssymb}
\usepackage{dcolumn}
\usepackage{bm}
\usepackage{color}
\usepackage{longtable}
\usepackage{nicefrac}
\usepackage{xfrac}
\newcommand{\cm}{\ensuremath{\textrm{cm\textsuperscript{-1}}}}
\newcommand{\cowan}{\mbox{\textsc{cowan}}}
\newcommand{\ambit}{\textsc{amb}{\footnotesize i}\textsc{t}}

\begin{document}
\title{Energy Level Structure of Sn$^{3+}$ Ions}

\author{J.~Scheers}
\affiliation{Advanced Research Center for Nanolithography, Science Park 110, 1098 XG Amsterdam, The Netherlands}
\affiliation{Department of Physics and Astronomy, and LaserLaB, Vrije Universiteit, De Boelelaan 1081, 1081 HV Amsterdam, The Netherlands}

\author{A.~Ryabtsev}
\affiliation{Institute of Spectroscopy, Russian Academy of Sciences, Troitsk, Moscow, 108840 Russia}

\author{A.~Borschevsky}
\affiliation{Van Swinderen Institute, University of Groningen, Nijenborgh 4, 9747 AG Groningen, The Netherlands}

\author{J.~C.~Berengut}
\affiliation{School of Physics, University of New South Wales, Sydney 2052, Australia}

\author{K.~Haris}
\affiliation{Department of Physics, Aligarh Muslim University, Aligarh, 202002, India}

\author{R.~Schupp}
\affiliation{Advanced Research Center for Nanolithography, Science Park 110, 1098 XG Amsterdam, The Netherlands}

\author{D.~Kurilovich}
\affiliation{Advanced Research Center for Nanolithography, Science Park 110, 1098 XG Amsterdam, The Netherlands}
\affiliation{Department of Physics and Astronomy, and LaserLaB, Vrije Universiteit, De Boelelaan 1081, 1081 HV Amsterdam, The Netherlands}

\author{F.~Torretti}
\affiliation{Advanced Research Center for Nanolithography, Science Park 110, 1098 XG Amsterdam, The Netherlands}
\affiliation{Department of Physics and Astronomy, and LaserLaB, Vrije Universiteit, De Boelelaan 1081, 1081 HV Amsterdam, The Netherlands}

\author{A.~Bayerle}
\affiliation{Advanced Research Center for Nanolithography, Science Park 110, 1098 XG Amsterdam, The Netherlands}

\author{E.~Eliav}
\affiliation{School of Chemistry, Tel Aviv University, 69978 Tel Aviv, Israel}

\author{W.~Ubachs}
\affiliation{Advanced Research Center for Nanolithography, Science Park 110, 1098 XG Amsterdam, The Netherlands}
\affiliation{Department of Physics and Astronomy, and LaserLaB, Vrije Universiteit, De Boelelaan 1081, 1081 HV Amsterdam, The Netherlands}

\author{O.~O.~Versolato}
\affiliation{Advanced Research Center for Nanolithography, Science Park 110, 1098 XG Amsterdam, The Netherlands}

\author{R.~Hoekstra}
\email[]{r.a.hoekstra@rug.nl}
\affiliation{Advanced Research Center for Nanolithography, Science Park 110, 1098 XG Amsterdam, The Netherlands}
\affiliation{Zernike Institute for Advanced Materials, University of Groningen, Nijenborgh 4, 9747 AG Groningen, The Netherlands}

\date{\today}

\begin{abstract}
Laser-produced Sn plasma sources are used to generate extreme ultraviolet (EUV) light in state-of-the-art nanolithography. An ultraviolet and optical spectrum is measured from a droplet-based laser-produced Sn plasma, with a spectrograph covering the range 200 - 800 nm. This spectrum contains hundreds of spectral lines from lowly charged tin ions Sn\textsuperscript{1+} - Sn\textsuperscript{4+} of which a major fraction was hitherto unidentified. We present and identify a selected class of lines belonging to the quasi-one-electron, Ag-like (\mbox{[Kr]$4d\textsuperscript{10}nl$} electronic configuration), Sn\textsuperscript{3+} ion, linking the optical lines to a specific charge state by means of a masking technique. These line identifications are made with iterative guidance from \cowan~code~calculations. Of the 53 lines attributed to Sn\textsuperscript{3+}, some 20 were identified from previously known energy levels, and 33 lines are used to determine previously unknown level energies of 13 electronic configurations, i.e.,  $ 7p $, $ (7,8)d $, $ (5,6)f $, $ (6-8)g $, $ (6-8)h $, $ (7,8)i $. The consistency of the level energy determination is verified by the quantum-defect scaling procedure. The ionization limit of Sn\textsuperscript{3+} is confirmed and refined to 328\,908.4\,\cm~with an uncertainty of 2.1\,\cm. The relativistic Fock space coupled cluster (FSCC) calculation of the measured level energies are generally in good agreement with experiment, but fail to reproduce the anomalous behavior of the \mbox{$5d\;^2$D} and \mbox{$nf\;^2$F} terms. By combining the strengths of FSCC, \cowan~code calculations, and configuration interaction many-body perturbation theory (CI+MBPT), this behavior is shown to arise from interactions with doubly-excited configurations.
\end{abstract}


\maketitle

\section{Introduction}
Emission of light by neutral tin atoms and lowly charged tin ions, \mbox{Sn\,\textsc{i}} - \mbox{Sn\,\textsc{v}}, is abundant in a wide variety of plasmas, ranging from laser-produced extreme-ultraviolet (EUV) light generating Sn plasma for nanolithography~\cite{Banine2011,Svendsen1994}, divertor plasma when using tin containing materials in future thermonuclear fusion reactors~\cite{Coenen2014,VanEden2016,VanEden2017}, discharge plasma between tin whiskers causing short-circuits \cite{Mason2007}, to astrophysical environments \cite{Hobbs1993,Proffitt2001,Chayer2005,Adelman1979,Savage1996,Sofia1999,Garcia2016,Milan2018,Biswas2018}. Spectroscopic investigations on these kinds of plasmas can help characterize plasma parameters \cite{Roy2015,Coons2010,Harilal2005,Lan2017,Iqbal2016,Namba2008,Kieft2004} such as ion and electron densities and temperatures by studying the observed line strengths and their shapes. However, spectroscopic information on the relevant charge states Sn\textsuperscript{3+} and Sn\textsuperscript{4+}, i.e., \mbox{Sn\,\textsc{iv}} and \mbox{Sn\,\textsc{v}}, is rather scarce, because of the poorly known electronic structure of these ions. 

Sn\textsuperscript{3+}, with its ground electronic configuration \mbox{[Kr]$4d^{10}\,5s$}, belongs to the Ag-like isoelectronic sequence. Remarkably, only the lowest eight singly-excited \mbox{$4d^{10}\,nl$}, the doubly-excited \mbox{$4d^{9}\,5s^2$} and three \mbox{$4d\textsuperscript{9}\,5s\,5p$} levels in Sn\textsuperscript{3+} are tabulated in the NIST database~\cite{NISTASD}. The level energies originate from unpublished work by Shenstone~\cite{Moore1958}, while wavelengths are given in another compilation by the National Bureau of Standards~\cite{Reader1980}. The assessment of energy levels by Shenstone is based on extended and revised work by Lang and others~\cite{Carroll1926,Rao1926,Gibbs1929,Lang1930}. Since the early compilation~\cite{Moore1958} of almost 60 years ago, the only extension of the electronic energy level structure of \mbox{Sn\,\textsc{iv}} stems from EUV spectroscopy by Ryabtsev and coworkers \cite{Ryabtsev2006} in which they extend the $ns$ series from $n=8$ up to $n=10$ and add the \mbox{$7d\;^2$D} term. A more extensive list of \mbox{Sn\,\textsc{iv}} lines is given in an otherwise unpublished MSc thesis~\cite{Wu1967}. In other works, beam-foil techniques have been used to determine lifetimes~\cite{Pinnington1987,Kernahan1985,Andersen1972}. Aside of the singly-excited levels, some doubly-excited energy levels belonging to the \mbox{$4d\textsuperscript{9}\,5s\,5p$} configuration are identified in laser- and vacuum-spark-produced tin plasmas~\cite{Kaufman1985,Dunne1992,Ryabtsev2006,Ryabtsev2007,Lysaght2005,Lysaght2005a}. Theoretical level energies and transition probabilities~\cite{Safronova2003,Ivanova2011} have been calculated for Ag-like ions. The narrow, inverted fine structure of the \mbox{$4f\;^2$F} term in Ag-like Sn\textsuperscript{3+} has been addressed in detail by theory~\cite{Safronova2003,Ivanova2011,Ding2012,Grumer2014,Cheng1979}. In spite of all these efforts, knowledge of the electronic structure of \mbox{Sn\,\textsc{iv}} is mostly limited to its lowest energy levels.

To obtain the electronic structure of Sn\textsuperscript{3+}, we have studied its line emission in the wavelength range of \mbox{200 -- 800\,nm}. The optical lines belonging to \mbox{Sn\,\textsc{iv}} are identified amongst the hundreds of optical lines stemming from a laser-produced droplet-based Sn plasma, by taking spectra as a function of laser intensity. The method to single out transitions belonging to ions in a specific charge state relies on the strongly changing ratio between line intensity and background emission from the plasma as a function of laser intensity. 

We will first introduce and detail a convenient method to obtain charge-state-resolved optical spectra from a laser-produced plasma (LPP) in the following. Out of the over 350 lines observed in the visible spectral range, 53 are identified as stemming from Sn\textsuperscript{3+}. Out of those, 33 lines are new determinations. Thereafter, the line identification is discussed. On basis of these line identifications an extended level diagram for Sn\textsuperscript{3+} is constructed. The consistency of the highly-excited levels is checked by quantum-defect scalings. In the final section, Fock space coupled cluster (FSCC) and configuration interaction many-body perturbation theory (CI+MBPT) calculations are employed to explain the anomalous behavior of the \mbox{$5d\;^2$D} and \mbox{$nf\;^2$F} terms. 

\section{Experimental setup} \label{sec:expsetup}
\begin{figure}[t]
	\includegraphics[width=\columnwidth]{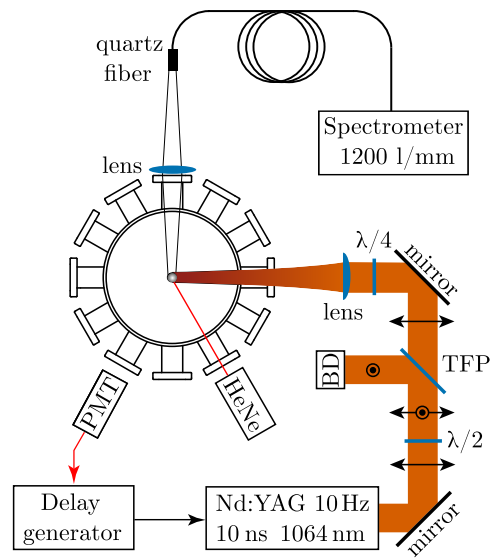}
	\caption{Schematic top view of the main components of the LPP source from which the spectroscopic data was taken. For details see Section~\ref{sec:expsetup}.  \label{experimentalsetup}}
\end{figure}

An overview of the experimental setup is depicted in Fig.~\ref{experimentalsetup}. A more detailed explanation is provided in Ref.~\cite{kurilovich2016}. The experimental LPP source consists of a vacuum vessel (about 10\textsuperscript{-7}\,mbar) equipped with a droplet generator from which a 10\,kHz stream of liquid tin micro-droplets is ejected. The droplets have a diameter of about 45\,$\mu$m. A 10-Hz pulsed Nd:YAG laser, operating at its fundamental wavelength of 1064\,nm, is used to irradiate the droplets in order to generate a plasma. The laser energy is varied without changing the beam shape by using the combination of a half-wave plate ($ \lambda /2 $) and a thin-film polarizer (TFP), reflecting part of the light into a beamdump (BD). The laser beam is  circularly polarized by a quarter-wave plate ($ \lambda /4 $), hereafter the beam is focused onto the droplet. This results in a Gaussian full-width-at-half-maximum (FWHM) beam size of 115\,$\mu$m at the droplet position. The laser has a 10\,ns FWHM pulse length. Light reflected by the droplet falling through a helium-neon (HeNe) laser sheet is detected by a photon-multiplier tube (PMT) used to trigger the laser. 

The light emitted from the plasma is observed through a viewport perpendicular to the laser beam propagation and 30 degrees above the horizontal plane. A biconvex lens images the plasma onto a quartz fiber that is used to guide the light to the spectrometer (Princeton Isoplane SCT 320). The entrance side of the fiber consists of 19~cores with a diameter of 200\,$\mu$m in a hexagonal configuration, while at the exit side the cores are oriented in a linear configuration to efficiently guide light through the spectrometer slit. The spectrometer is laid out in a Czerny-Turner configuration with a focal length of 320\,mm. The grating has 1200 lines per mm and is blazed at 500\,nm, leading to a significantly reduced grating diffraction efficiency below 300\,nm. A CCD camera (Princeton Pixis 2KBUV) optimized for the ultraviolet and visible regime recorded the diffracted light. By rotating the grating, thus changing the spectral detection range, the full spectral range from 200 to 800\,nm is covered in steps of approximately 50\,nm, overlapping by about 10\,nm. From the shortest to the longest wavelength the linear dispersion decreases from 0.033 to 0.028\,nm per pixel. 

The wavelength axis is calibrated using neon-argon and mercury lamps. The FWHM line widths of the calibration lines are smaller than 0.1\,nm. The total uncertainties of the mid positions of the Sn\textsuperscript{3+} lines are better than 0.01\,nm over all observed laser energies and wavelengths. The emitted light is space- and time-integrated by summing the intensity resulting from the various fiber cores and taking an integration time of 10\,s, corresponding to 100 laser shots.

Measurements are performed with, and without, edge-pass filters to distinguish second-order lines from first-order ones. This enables filtering out the second-order lines appearing at wavelengths longer than 400\,nm. 
Additionally, closely packed lines in the ultraviolet below 300\,nm can be determined in second order at a higher resolution. Weakly appearing lines in first order, due to the low grating response below 300\,nm, are observed with a higher intensity in second order.

\begin{figure*}[t]
	\includegraphics[width=\textwidth]{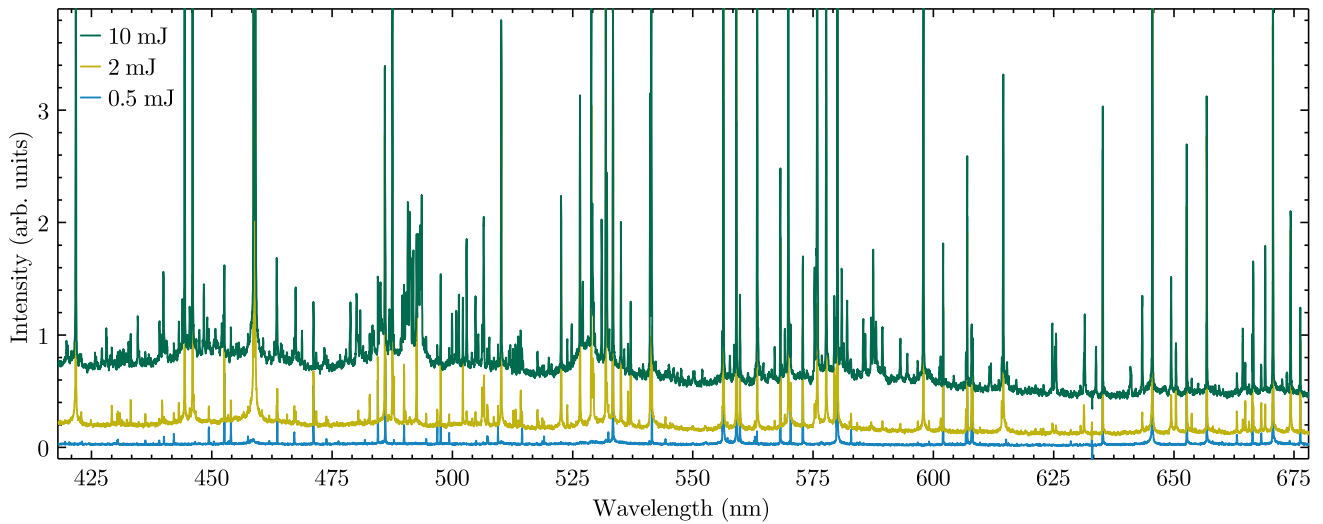}
	\caption{Experimentally obtained Sn\textsuperscript{q+} spectra for laser energies of 0.5\,mJ (blue, lower), 2\,mJ (yellow, middle), and 10\,mJ (green, upper). The observable increase in the background level is due to increased continuum emission from the plasma for higher laser energies. The spectra shown are taken without spectral filters and, thus, include second order contributions.\label{longgraph}}
\end{figure*}

\section{Charge state identification}\label{sec:csi}
We performed passive spectroscopy measurements on the laser-produced tin plasma for a series of laser energies ranging from 0.5 to 370\,mJ. Fig.~\ref{longgraph} shows example measurements over a selected wavelength range for three laser energies, where it is shown that the number of lines increases with laser energy. This is a signature of an increasing number of contributing charge states to the measured spectrum. A closer inspection indicates that indeed sets of lines appear with increasing laser energy that exhibit similar changes in intensity. As will be demonstrated below, each of these sets of lines can be singled out by considering their intensities with respect to the continuum background, increasing strongly with laser energy. 

\begin{figure*}
	\includegraphics[width=\textwidth]{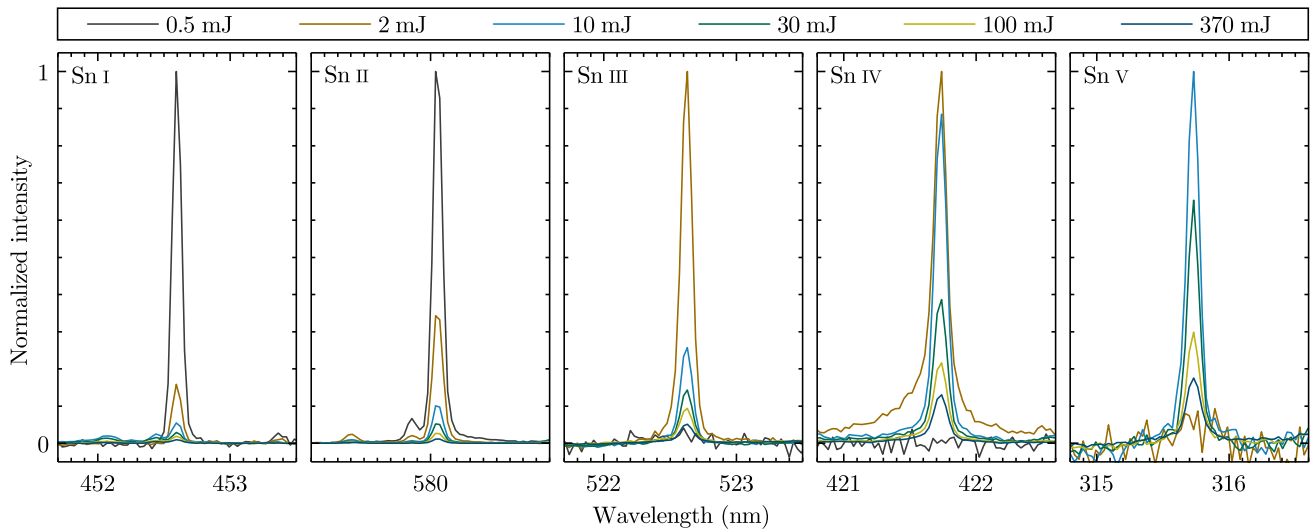}
	\caption{Spectral intensity scaling as a function of wavelength (vacuum, in nm) for varying laser energy. Here, the intensity is normalized to their (local) continuum background level, and unity is subsequently subtracted. The laser energies depicted are 0.5\,mJ (black), 2\,mJ (brown), 10\,mJ (light blue), 30\,mJ (green), 100\,mJ (yellow) and 370\,mJ (dark blue). The specific transitions shown are described in the text. \label{scaling}}
\end{figure*}

To illustrate the procedure we select a well-known line of each of the charge states \mbox{Sn\,\textsc{i}}$-$\mbox{Sn\,\textsc{v}}. These are the \mbox{Sn\,\textsc{i}} \mbox{$5p^{2}\;^{1}\textrm{S}_{0} - 5p\,6s\; ^{1}\textrm{P}_{1}$} \mbox{($\lambda$=452.60\,nm~\cite{Meggers1940})}, \mbox{Sn\,\textsc{ii}} \mbox{$5d\; ^{2} \textrm{D} _{\textrm{\sfrac{5}{2}}} - 4f\; ^{2}\textrm{F}_{\textrm{\sfrac{7}{2}}} $} \mbox{($\lambda$=580.05\,nm~\cite{Haris2014})} (the transition from the \mbox{$4f\; ^{2}\textrm{F}_{\textrm{\sfrac{5}{2}}} $} at 579.85\,nm is also visible), \mbox{Sn\,\textsc{iii}} \mbox{$6s\;^{1}\textrm{S}_{0} - 6p\;^{1}\textrm{P}_{1} $} \mbox{($\lambda$=522.64\,nm~\cite{Haris2012})}, \mbox{Sn\,\textsc{iv}} \mbox{$6s\; ^{2}\textrm{S}_{\textrm{\sfrac{1}{2}}} - 6p\; ^{2}\textrm{P}_{\textrm{\sfrac{1}{2}}} $ } \mbox{($\lambda$=421.73\,nm)} and \mbox{Sn\,\textsc{v}} \mbox{$6s\; ^{3}\textrm{D}_{3} - 6p\; ^{3}\textrm{F}_{4} $ } ($\lambda$=315.6\,nm, based on level energies taken from Refs.~\cite{vanKleef1981,Ryabtsev2005}). The scaled intensity is defined as $\sfrac{I_{\lambda}}{I\textsubscript{bg}} - 1$, with $I_{\lambda}$ the line intensity and $I\textsubscript{bg}$ the (local value of the) continuum background level. For direct comparison, the scaled intensity of each individual line is normalized to its maximum value, as shown in Fig.~\ref{scaling}.
From Fig.~\ref{scaling}, it is seen that for the lowest charge states \mbox{Sn\,\textsc{i}} and \mbox{Sn\,\textsc{ii}} the normalized, scaled intensity maximizes for the lowest laser energy of 0.5\,mJ, while for the highest observed charge state \mbox{Sn\,\textsc{v}} a laser energy of 10\,mJ is optimal. \mbox{Sn\,\textsc{iv}}, the ion of interest here, maximizes at 2\,mJ. This demonstrates that the contributions of higher charge states to the spectrum increase with increasing laser energy. Preliminary nanosecond time-resolved spectroscopic measurements revealed that spectral line emission is mostly observed in the late-time evolution of the plasma. Traces of line broadening are observed in the time-integrated spectra presented in this work, e.g., the \mbox{Sn\,\textsc{iv}} line shape in the 2\,mJ spectrum. Analysis of line broadening mechanisms and the time-evolution of these plasmas will be left for future work as they do not influence our line identifications and are outside the scope of this paper.

\begin{figure}[]
	\includegraphics[width=\columnwidth]{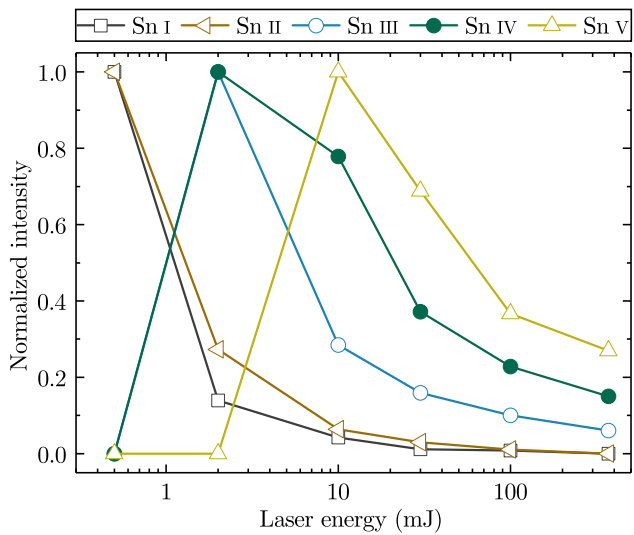}
	\caption{Normalized scaled intensities of \mbox{Sn\,\textsc{i}}--\mbox{Sn\,\textsc{v}} lines as a function of laser energy. The same lines as presented in Fig.~\ref{scaling} are used. Other \mbox{Sn\,\textsc{i}}--\mbox{Sn\,\textsc{v}} lines show a similar dependence on laser energy. \label{normalizedplot}}
\end{figure}

Fig.~\ref{normalizedplot} quantifies the dependence of the scaled intensities for spectral lines belonging to tin ions in charge states 0, 1, 2, 3, and 4+ produced in the Sn LPP. The unique energy dependence of each charge state enables a straightforward assignment of unknown lines to specific charge states. In this way 53 lines are assigned to \mbox{Sn\,\textsc{iv}}.

\section{Line identification procedure}
Of the 53 lines in the ultraviolet and optical spectral range attributed to \mbox{Sn\,\textsc{iv}}, 20 are readily identified as transitions between energy levels known from literature~\cite{Moore1958}, and match well within mutual experimental uncertainties with the line positions given in Ref.~\cite{Wu1967}. These 20 lines are presented in Table \ref{tab:firstlines} along with their connecting upper and lower energy levels. 

Having identified 20 lines using known energy levels, we proceed with the identification of the other lines ascribed to \mbox{Sn\,\textsc{iv}}.  Unique identifications of the observed lines requires an accuracy of the level energies of better than 10\textsuperscript{-3}, which is challenging for atomic theories. Therefore, an iterative procedure is used to identify the unknown lines. We use the \cowan~code to calculate the electronic structure and transitions, and adjust its parameters to match perfectly the known lines in the spectrum. In this way, energy levels just above the known ones can be obtained with sufficient accuracy to identify a next set of lines. This procedure can be repeated to identify all lines. Furthermore, quantum defect theory~\cite{Edlen1964}~is used to check the consistency of level energies for each $l$ series.

\begin{table}
	\caption{Vacuum wavelengths (in nm) of \mbox{Sn\,\textsc{iv}} lines between levels previously known. The wavelengths determined in this work are compared with literature values taken from an otherwise unpublished Master's thesis~\cite{Wu1967}. The upper and lower energy levels indicating the transition represent the $ nl $ one-electron orbital outside the \mbox{[Kr]$4d^{10}$} core configuration. The $ 5s $\textsuperscript{2} indicates the doubly-excited \mbox{$4d^{9}\,5s^2$} configuration. The wavelengths determined in this work are averaged centroid positions of Gaussian fits in spectra taken at different laser energies. The intensity $ I $ represents the area-under-the-curve of this line in the 30\,mJ spectrum.  $ gA $ factors from the upper level result from analysis with the \cowan~code. \label{tab:firstlines}}
	\begin{ruledtabular}
		\begin{tabular}{c|ccccccc}
			\multicolumn{2}{c}{$\lambda$\,(nm)} &                        $ I $                         &              $ gA $               &         \multicolumn{2}{c}{lower}         & \multicolumn{2}{c}{upper} \\ \cline{1-2}\cline{5-6}\cline{7-8}
			this work &       literature        &                      (arb.un.)                       & ($ 10^8 $\,s\textsuperscript{-1}) &          $nl$           &      $ J $      & $nl$ &       $ J $        \\ \hline
			 208.23   &         208.224         &                    \hphantom{0}25                    &              130.2\hphantom{1}               &          $4f$           & \nicefrac{7}{2} & $5g$ &  \nicefrac{9}{2}   \\
			 208.49   &         208.485         &                    \hphantom{0}20                    &              100.4\hphantom{1}                &          $4f$           & \nicefrac{5}{2} & $5g$ &  \nicefrac{7}{2}   \\
			 222.14   &         222.156         &                         289                          &         53.8         &          $5d$           & \nicefrac{3}{2} & $4f$ &  \nicefrac{5}{2}   \\
			 222.66   &         222.680         &                    \hphantom{00}1                    &         \hphantom{0}3.4         &          $5d$           & \nicefrac{5}{2} & $4f$ &  \nicefrac{5}{2}   \\
			 222.96   &         222.980         &                    \hphantom{0}34                    &         68.5         &          $5d$           & \nicefrac{5}{2} & $4f$ &  \nicefrac{7}{2}   \\
			 243.75   &         243.757         &                    \hphantom{0}35                    &         \hphantom{3}6.6         & \hphantom{\textsuperscript{2}}$5s$\textsuperscript{2} & \nicefrac{5}{2} & $4f$ &  \nicefrac{7}{2}   \\
			 251.47   &         251.466         &                    \hphantom{0}22                    &         \hphantom{3}6.0         &          $6p$           & \nicefrac{1}{2} & $7s$ &  \nicefrac{1}{2}   \\
			 266.05   &         266.054         & \hphantom{0}48 &         10.1         &          $6p$           & \nicefrac{3}{2} & $7s$ &  \nicefrac{1}{2}   \\
			 270.68   &         270.667         &                         154                          &         27.5         &          $6p$           & \nicefrac{1}{2} & $6d$ &  \nicefrac{3}{2}   \\
			 284.91   &         284.922         &                         351                          &         42.2         &          $6p$           & \nicefrac{3}{2} & $6d$ &  \nicefrac{5}{2}   \\
			 287.64   &         287.633         &                    \hphantom{0}38                    &         \hphantom{3}4.6         &          $6p$           & \nicefrac{3}{2} & $6d$ &  \nicefrac{3}{2}   \\
			 287.96   &         287.961         &                    \hphantom{0}32                    &         \hphantom{3}1.5         &          $5d$           & \nicefrac{3}{2} & $6p$ &  \nicefrac{3}{2}   \\
			 288.85   &         288.840         &                         219                          &         11.2         &          $5d$           & \nicefrac{5}{2} & $6p$ &  \nicefrac{3}{2}   \\
			 307.26   &         307.247         &                         251                          &         \hphantom{3}6.0         &          $5d$           & \nicefrac{3}{2} & $6p$ &  \nicefrac{1}{2}   \\
			 324.71   &         324.700         &                         166                          &         \hphantom{3}1.6         & \hphantom{\textsuperscript{2}}$5s$\textsuperscript{2} & \nicefrac{5}{2} & $6p$ &  \nicefrac{3}{2}   \\
			 386.23   &         386.232         &                 1\,063\hphantom{1\,}                 &         \hphantom{0}9.2         &          $6s$           & \nicefrac{1}{2} & $6p$ &  \nicefrac{3}{2}   \\
			 402.08   &         402.071         &                         289                          &         \hphantom{3}4.5         &          $4f$           & \nicefrac{7}{2} & $6d$ &  \nicefrac{5}{2}   \\
			 403.03   &         403.076         &                    \hphantom{0}30                    &         \hphantom{3}0.2         &          $4f$           & \nicefrac{5}{2} & $6d$ &  \nicefrac{5}{2}   \\
			 408.52   &         408.520         &                         219                          &         \hphantom{3}3.0         &          $4f$           & \nicefrac{5}{2} & $6d$ &  \nicefrac{3}{2}   \\
			 421.73   &         421.735         &                         677                          &         \hphantom{3}3.5         &          $6s$           & \nicefrac{1}{2} & $6p$ &  \nicefrac{1}{2}
		\end{tabular}
	\end{ruledtabular}
\end{table}

\subsection{COWAN procedure}
The \cowan~code~\cite{Cowan1981}, one of the most widely applied electronic structure codes, is used to calculate the energies of yet unestablished \mbox{Sn\,\textsc{iv}} levels. The \cowan~code produces radial wave functions using a quasi-relativistic Hartree-Fock method. The electrostatic single configuration radial integrals F\textsubscript{k} and G\textsubscript{k} (Slater integrals), configuration interaction, Coulomb radial integrals and spin-orbit parameters are calculated from the obtained wavefunctions. Subsequently, level energies and intermediate coupling eigenvectors are extracted. Furthermore, values for the transition probabilities and wavelengths are obtained. 

The values of the electrostatic integrals are scaled by a factor between 0.7 and 0.85 as well as the scaling of spin-orbit parameters to optimally fit the thus far understood experimental spectrum. The outcome of this parameter scaling procedure yields a useful interpretation of the experimental spectrum. This enables predictions for lines between not yet experimentally established energy levels. These predictions include their relative strength expressed as the $gA$ factor: the Einstein coefficient $A$ multiplied with the statistical weight $g$ of the upper state. Due to the one-electron nature of the system, the number of allowed transitions between two terms is only three (or two if a $^2$S term is involved). However, we might observe only two lines since transitions between equal angular momenta have typically a small $ gA $ factor. Experimental lines that lie close to a predicted transition and have relative line strengths similar to the ones determined on basis of the aforementioned $gA$ factors are thus assigned to a specific transition. This provides an enlarged set of levels that can be used to fine-tune the calculations in a next step.

\subsection{Quantum defect}
The energy levels of quasi-one-electron systems approach a hydrogen-like level structure, especially for high principal and angular quantum numbers. For such systems the energy levels which are shifted towards slightly higher binding energies can be well-described by introducing the so-called quantum defect $\delta_{l}$ as a correction to the Bohr formula. The position of energy level $E_{nl}$ (relative to the ionization limit) is defined by~\cite{Edlen1964}

\begin{equation}
E_{nl}=-R\frac{Z_c^{2}}{(n-\delta_{l})^{2}}, 
\label{eq:qdformula}
\end{equation}

\noindent
with $ Z_c $ the net charge state of the core ($ Z_c =4$ for Sn\textsuperscript{3+}), and  $ n $ the principal quantum number. $ R $ relates to the Rydberg constant $ R_{\infty} $ as $R=R_{\infty}(1+\nicefrac{m_e}{M})^{-1}$, with $ m_e $ and $ M $ the electronic and nuclear mass, respectively. 
Following the review by Edl\'en~\cite{Edlen1964}, the quantum defect can be written as a Taylor expansion in $1/n^{*^2}$ with $ n^* $ being the apparent principal quantum number $ n^* = n - \delta_{l} $, with quantum defect $ \delta_{l} $,

\begin{equation}
\delta_l = a \left( \frac{1}{n^{*^2}} \right)+b\left(\frac{1}{n^{*^2}}\right)^2 + \dots
\label{eq:qd}
\end{equation}

The quantum defect becomes smaller with increasing angular momentum. For high-$ n $ values the first term dominates and the minute change of  $\delta_{l}$ as a function of  $1/n^{*^2}$ becomes linear. Additionally, it is well-established that $ a $  is positive for $ l \leq l$\textsubscript{core}, while $ a $ is negative for $ l > l$\textsubscript{core}~\cite{Edlen1964}. For Sn\textsuperscript{3+} with its $4d$\textsuperscript{10} core, $l$\textsubscript{core}$=2$.

\section{Results and discussion}
Using the iterative guidance from the \cowan~code as described above, we assign the newly found lines. The new \mbox{Sn\,\textsc{iv}} line assignments are summarized in Table~\ref{tab:wavelength}. The level energies of excited states are determined with respect to the $ 6s $, which is used as a anchor level considering that transitions to the $ 5s $ ground state are outside our detection region. The optimization of level energies is performed using Kramida's code~\textsc{lopt}~\cite{Kramida2011}, and the final results are presented in Table~\ref{tab:energylevels}.

\begin{table}[h!]
	\caption{Assignments and vacuum wavelengths (in nm) of UV and visible transitions of \mbox{Sn\,\textsc{iv}} identified in this work. The upper and lower energy levels indicating the transition represent the $ nl $ one-electron orbital outside the \mbox{[Kr]$4d^{10}$} core configuration. The wavelengths determined in this work are averaged centroid positions of Gaussian fits in spectra taken at different laser energies. The intensity $ I $ represents the area-under-the-curve of this line in the 30\,mJ spectrum. $ gA $ factors from the upper level result from analysis with the \cowan~code. The fine structure for several high-$ l $ states could not be resolved experimentally, therefore no individual angular momenta are given and the reported $ gA $ is the summed value of the three possible transitions. 
		\label{tab:wavelength}}
	\begin{ruledtabular}
		\begin{tabular}{ccccccc}
			              $\lambda$               &                          $ I $                           &             $ gA $             &    \multicolumn{2}{c}{lower}    & \multicolumn{2}{c}{upper} \\ \cline{4-5}\cline{6-7}
			                (nm)                  &                        (arb.un.)                         & ($10^8$\,s\textsuperscript{-1}) & $nl$ &          $ J $           & $ nl $ &      $ J $       \\ \hline
			               202.08                 &                      \hphantom{50}1                      &             10.5              & $5f$ &     \nicefrac{7}{2}      &  $8g$  &         \nicefrac{9}{2}         \\
			               203.36                 &                      \hphantom{50}2                      &        \hphantom{1}8.6        & $5f$ &     \nicefrac{5}{2}      &  $8g$  &           \nicefrac{7}{2}       \\
			               231.72                 &                      \hphantom{50}2                      &            14.4             & $5g$ &           &  $8h$  &                  \\
			               242.55                 &                      \hphantom{50}3                      &        \hphantom{1}0.3       & $6d$ &     \nicefrac{3}{2}      &  $6f$  & \nicefrac{5}{2}  \\
			               245.15                 &                      \hphantom{50}1                      &        \hphantom{1}0.5        & $6d$ &     \nicefrac{5}{2}      &  $6f$  & \nicefrac{7}{2}  \\
			               273.75                 &                      \hphantom{50}3                      &        \hphantom{1}2.2        & $7p$ &     \nicefrac{1}{2}      &  $8d$  & \nicefrac{3}{2}  \\
			               279.91                 &                      \hphantom{50}8                      &        \hphantom{1}3.8       & $7p$ &     \nicefrac{3}{2}      &  $8d$  & \nicefrac{5}{2}  \\
			               287.82                 &                      \hphantom{5}48                      &             30.4              & $5g$ &          &  $7h$  &                  \\
			               357.01                 &                           204                            &             24.6              & $5f$ &     \nicefrac{7}{2}      &  $6g$  &    \nicefrac{9}{2}           \\
			               360.99                 &                           233                            &             20.2              & $5f$ &     \nicefrac{5}{2}      &  $6g$  & \nicefrac{7}{2}  \\
			               393.41                 &                      \hphantom{5}27                      &        \hphantom{1}5.7        & $6f$ &     \nicefrac{7}{2}      &  $8g$  &           \nicefrac{9}{2}       \\
			               395.08                 &  \hphantom{1}35  &        \hphantom{1}4.4        & $6f$ &     \nicefrac{5}{2}      &  $8g$  & \nicefrac{7}{2}                 \\
			               459.04                 &                   2\,518\hphantom{2\,}                   &             93.7              & $5g$ &          &  $6h$  &                  \\
			               463.49                 &                           202                            &             15.1              & $6g$ &       &  $8h$  &                  \\
			               467.41                 &                           107                            &             14.7              & $6h$ &                          &  $8i$  &                  \\
			               504.82                 &                      \hphantom{5}90                      &        \hphantom{1}2.0        & $7p$ &     \nicefrac{1}{2}      &  $8s$  & \nicefrac{1}{2}  \\
			               529.12                 &  \hphantom{5}40  &        \hphantom{1}3.5        & $7p$ &     \nicefrac{3}{2}      &  $8s$  & \nicefrac{1}{2}  \\
			               541.12                 &                           366                            &        \hphantom{1}7.9        & $7p$ &     \nicefrac{1}{2}      &  $7d$  & \nicefrac{3}{2}  \\
			               563.38                 &                           662                            &             12.4              & $7p$ &     \nicefrac{3}{2}      &  $7d$  & \nicefrac{5}{2}  \\
			               563.60                 &  \hphantom{5}60 &        \hphantom{1}0.7        & $5g$ &     \nicefrac{7}{2}      &  $6f$  & \nicefrac{5}{2}  \\
			               567.00                 &                      \hphantom{5}42                      &        \hphantom{1}0.9        & $5g$ &     \nicefrac{9}{2}      &  $6f$  & \nicefrac{7}{2}  \\
			               569.13                 &                      \hphantom{5}68                      &        \hphantom{1}1.3        & $7p$ &     \nicefrac{3}{2}      &  $7d$  & \nicefrac{3}{2}  \\
			               575.85                 &                           881                            &             10.0              & $6d$ &     \nicefrac{3}{2}      &  $5f$  & \nicefrac{5}{2}  \\
			               589.40                 &                      \hphantom{5}71                      &        \hphantom{1}6.9        & $6f$ &     \nicefrac{7}{2}      &  $7g$  &       \nicefrac{9}{2}           \\
			               593.13                 &                           368                            &        \hphantom{1}5.3        & $6f$ &     \nicefrac{5}{2}      &  $7g$  &        \nicefrac{7}{2}          \\
			               597.93                 &                           940                            &             12.5              & $6d$ &     \nicefrac{5}{2}      &  $5f$  & \nicefrac{7}{2}  \\
			               643.43                 &                           154                            &        \hphantom{1}3.3        & $5f$ &     \nicefrac{7}{2}      &  $7d$  & \nicefrac{5}{2}  \\
			               664.32                 &                           103                            &        \hphantom{1}2.3        & $5f$ &     \nicefrac{5}{2}      &  $7d$  & \nicefrac{3}{2}  \\
			               673.51                 &                      \hphantom{5}36                      &        \hphantom{1}0.5        & $6d$ &     \nicefrac{3}{2}      &  $7p$  & \nicefrac{3}{2}  \\
			               688.89                 &                           212                            &        \hphantom{1}3.8        & $6d$ &     \nicefrac{5}{2}      &  $7p$  & \nicefrac{3}{2}  \\
			               717.44                 &                      \hphantom{5}98                      &        \hphantom{1}1.8        & $6d$ &     \nicefrac{3}{2}      &  $7p$  & \nicefrac{1}{2}  \\
			               759.65               & 316 &             49.6              & $6g$ &           &  $7h$  &                  \\
			               769.75               & 388 &             49.6              & $6h$ &                          &  $7i$  &
		\end{tabular}
	\end{ruledtabular}
\end{table}

The consistency of the energies of levels within a specific $l$ series is verified by determining the respective quantum defects. Quantum defects are calculated using Eq.~(\ref{eq:qdformula}) with the found level energies relative to the ionization limit. Therefore an accurate value of this limit is needed. 

\begin{figure}
	\includegraphics[width=\columnwidth]{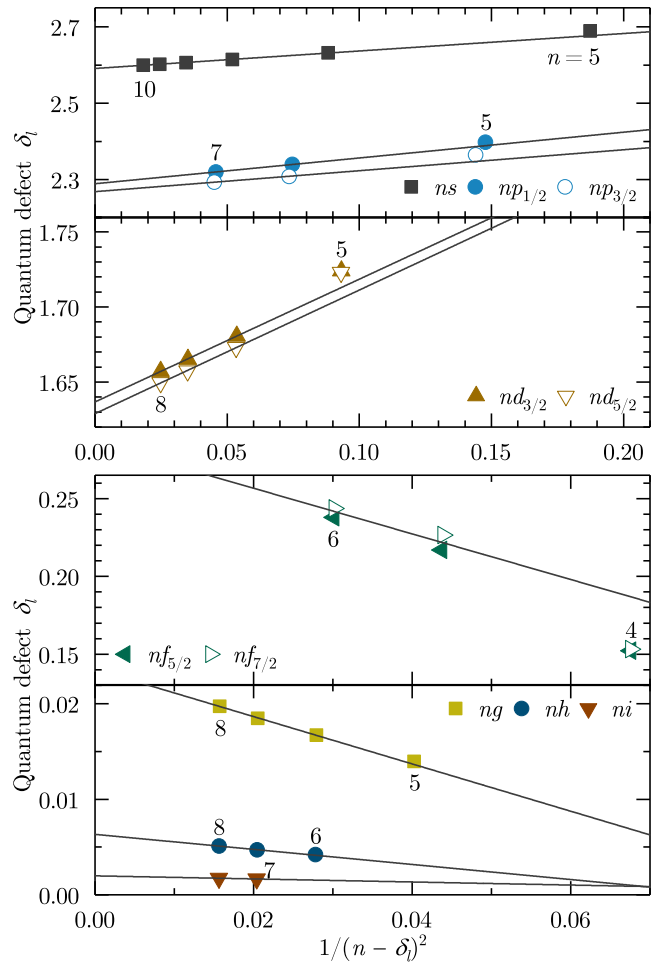}
	\label{fig:quantumdefects}
	\caption{Quantum defect values as a function of $1/n^{*^2}$ of the \mbox{Sn\,\textsc{iv}} energy levels, for $ l \leq l$\textsubscript{core} (upper graph) and $ l > l$\textsubscript{core} (lower graph). The quantum defects are calculated using Eq.~(\ref{eq:qdformula}) and the refined ionization limit of 328\,908.4\,\cm. The black lines are linear fits of the data points where the lowest level is excluded (except for the $ ni $ configuration).  \label{quantumdefects}}
\end{figure}

The ionization limit of 328\,550\,(300)\,\cm, tabulated in the NIST database~\cite{NISTASD}, is based on the determination of the series limit of the $ns$ levels ($n=5-7$). Ryabtsev~\textit{et~al.}~\cite{Ryabtsev2006} extended this $ns$ series with $8s$, $9s$, and $10s$. Using the extended $ns$ series they were able to refine the ionization limit to 328\,910\,(5)\,\cm. The ionization limit can be further improved by taking additional levels into account. Configurations which are prone to shifting of the level energies by configuration interaction effects ($ np $, $ nd $, and $ nf $, further described in Section~\ref{sec:fseffects}) are however unsuitable for determining the series limit. Therefore we only use the $ ng $, $ nh $ and $ ni $ configurations to refine the ionization limit. The results from analysis with the \textsc{polar} code~\cite{Polar}~is 328\,908.4\,\cm~with a statistical error of 0.3\,\cm. Combining this in quadrature with the uncertainty of $ 6s $ anchor level, we arrive at a total uncertainty of 2.1\,\cm. Fig.~\ref{quantumdefects} presents the quantum defects of the \mbox{Sn\,\textsc{iv}} levels as a function of $1/n^{*^2}$. Overall, a smooth dependence is found for all angular quantum numbers from $l$=0 ($s$) up to $l$=6 ($i$), underpinning the consistency of our identifications. 

\begin{figure}
	\includegraphics[width=\columnwidth]{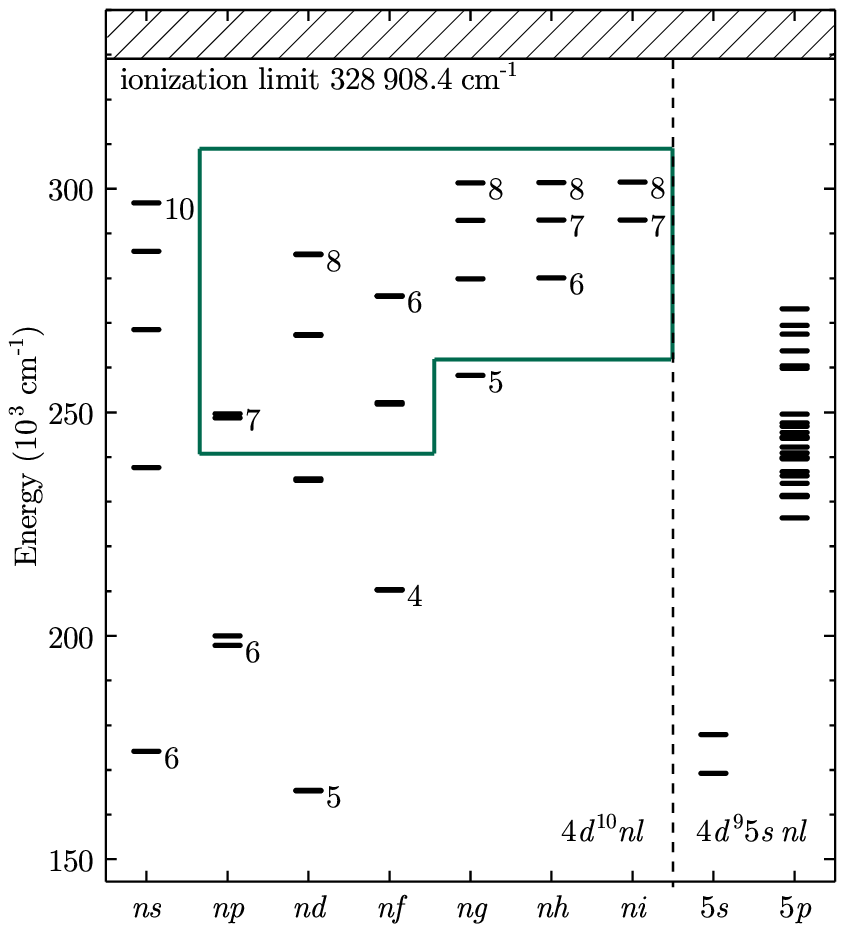}
	\caption{Level diagram of Sn\textsuperscript{3+}, drawn from 150\,000\,\cm~to the ionization limit at 328\,908.4\,\cm. The ground state $ 5s $, the $ 5p\;^2$P term, and multiply excited configurations lying near the ionization limit are omitted as transitions to these levels occur outside the detection range of this study. The levels determined by this study are shown in the boxed area. The other levels are based on Refs.~\cite{Moore1958,Ryabtsev2006}. \label{fig:leveldiagram}} 
\end{figure}

For the discussion of details of the line assignments and energy levels we consider separately levels for which the valence electron does or does not penetrate the electronic core, i.e., levels with $ l \leq l$\textsubscript{core} and $l > l$\textsubscript{core}, respectively. For Sn\textsuperscript{3+} with its $4d$\textsuperscript{10} core, $l$\textsubscript{core}$=2$. Anomalous effects on the fine structure splitting of the $ 5d $ and $ nf $ configurations are discussed and explained separately. Finally, Fig.~\ref{fig:leveldiagram} depicts the extended level diagram of \mbox{Sn\,\textsc{iv}} as a concise summary of our results.

\begin{table*}
	\caption{Energy levels of Sn\textsuperscript{3+}, with its ground state \mbox{[Kr]$4d^{10}\,5s$}. The experimental values obtained in this work are presented, next to the known values from literature given in Refs.~\cite{Moore1958,Ryabtsev2006}. The experimental level energies of excited states are calculated with respect to the $ 6s $ anchor level and are the result from analysis with the \textsc{lopt}~code. The statistical uncertainty is presented in parentheses. The Total column contains the sum of FSCC calculations including Breit interaction and QED effects. As a comparison to theoretical values, relativistic many-body perturbation theory (RMBPT) calculations obtained from Ref.~\cite{Safronova2003} are shown, while other known fine structure splittings are given in Table~\ref{tab:fstabel}. The fine structure splitting of several high-$ nl $ levels are smaller than 0.5\,\cm~and not resolved experimentally. In those cases, a value of the angular momentum is omitted. The ionization potential (IP) is presented at the bottom of the Table. \label{tab:energylevels}}
	\begin{ruledtabular}
		\begin{tabular}{ccx{2cm}|x{2cm}x{2cm}x{1.2cm}x{1.2cm}x{2cm}|x{2cm}}
			                                                                                                                     &                 &       \multicolumn{2}{c}{$E$\textsubscript{experiment} (\cm)}       &                                           \multicolumn{5}{c}{$E$\textsubscript{theory} (\cm)}                                            \tn \cline{3-4}\cline{5-9}
			                                                        $nl$                                                         &      $ J $      &        this work        &        literature\hphantom{~[20]}         &        FSCC         & $\Delta E$\textsubscript{Breit} & $\Delta E$\textsubscript{QED} &        Total        & RMBPT~\cite{Safronova2003} \tn \hline
			                                                        $5p$                                                         & \nicefrac{1}{2} &                         &  \hphantom{1}69\,563.9~\cite{Moore1958}   & \hphantom{1}69\,850 &         \hphantom{-1}62         &             -171              & \hphantom{1}69\,741 &    \hphantom{1}69\,265     \tn
			                                                                                                                     & \nicefrac{3}{2} &                         &  \hphantom{1}76\,072.3~\cite{Moore1958}   & \hphantom{1}76\,447 &         \hphantom{1}-26         &             -165              & \hphantom{1}76\,256 &    \hphantom{1}75\,736     \tn
			                                                        $5d$                                                         & \nicefrac{3}{2} &        165\,304(1)\hphantom{.1}         &        165\,304.7~\cite{Moore1958}        &      165\,974       &              -123               &             -205              &      165\,646       &          164\,538          \tn
			                                                                                                                     & \nicefrac{5}{2} &        165\,409(1)\hphantom{.1}         &        165\,410.8~\cite{Moore1958}        &      166\,731       &              -145               &             -204              &      166\,382       &          165\,283          \tn
			\hphantom{\textsuperscript{2}}$4d$\textsuperscript{9}$\, 5s $\textsuperscript{2}\hphantom{$4d$\textsuperscript{9}\,} & \nicefrac{5}{2} &        169\,233.6(8)         &        169\,233.6~\cite{Moore1958}        &                     &                                 &                               &                     &                            \tn
			                                                                                                                     & \nicefrac{3}{2} &                         &        177\,889.0~\cite{Moore1958}        &                     &                                 &                               &                     &                            \tn
			                                                        $6s$                                                         & \nicefrac{1}{2} &        174\,138.8(4)         &        174\,138.8~\cite{Moore1958}        &      174\,478       &         \hphantom{1}-99         &             -143              &      174\,236       &                            \tn
			                                                        $6p$                                                         & \nicefrac{1}{2} &        197\,850.6(6)         &        197\,850.9~\cite{Moore1958}        &      198\,292       &         \hphantom{1}-74         &             -193              &      198\,025       &                            \tn
			                                                                                                                     & \nicefrac{3}{2} &        200\,030.1(4)         &        200\,030.8~\cite{Moore1958}        &      200\,512       &              -103               &             -193              &      200\,216       &                            \tn
			                                                        $4f$                                                         & \nicefrac{7}{2} &        210\,258.2(6)         &        210\,257.7~\cite{Moore1958}        &      210\,912       &              -158               &             -200              &      210\,554       &          209\,418          \tn
			                                                                                                                     & \nicefrac{5}{2} &        210\,317.9(7)         &        210\,318.2~\cite{Moore1958}        &      210\,983       &              -156               &             -200              &      210\,627       &          209\,494          \tn
			                                                        $6d$                                                         & \nicefrac{3}{2} &        234\,797.0(1)         &        234\,795.7~\cite{Moore1958}        &      235\,509       &              -134               &             -203              &      235\,171       &                            \tn
			                                                                                                                     & \nicefrac{5}{2} &        235\,128.7(2)         &        235\,127.7~\cite{Moore1958}        &      235\,842       &              -144               &             -201              &      235\,497       &                            \tn
			                                                        $7s$                                                         & \nicefrac{1}{2} &        237\,617(1)\hphantom{.1}         &        237\,615.7~\cite{Moore1958}        &      238\,219       &              -123               &             -175              &      237\,920       &                            \tn
			                                                        $7p$                                                         & \nicefrac{1}{2} &        248\,735.4(2)         &                                           &      249\,402       &              -110               &             -197              &      249\,094       &                            \tn
			                                                                                                                     & \nicefrac{3}{2} &        249\,644.8(1)         &                                           &      250\,454       &              -124               &        \hphantom{1}-97        &      250\,233       &                            \tn
			                                                        $5f$                                                         & \nicefrac{7}{2} &        251\,853.0(2)         &                                           &      252\,984       &              -157               &             -201              &      252\,626       &          250\,981          \tn
			                                                                                                                     & \nicefrac{5}{2} &        252\,162.6(2)         &                                           &      253\,023       &              -155               &             -202              &      252\,666       &          251\,025          \tn
			                                                        $5g$                                                         & \nicefrac{7}{2} &        258\,283.2(3)         &        258\,282.3~\cite{Moore1958}        &      258\,782       &              -143               &             -201              &      258\,439       &          256\,868          \tn
			                                                                                                                     & \nicefrac{9}{2} &        258\,283.2(3)         &        258\,282.7~\cite{Moore1958}        &      258\,782       &              -143               &             -201              &      258\,439       &          256\,872          \tn
			                                                        $7d$                                                         & \nicefrac{3}{2} &        267\,215.5(2)         &      267\,247.6~\cite{Ryabtsev2006}       &      267\,815       &              -138               &             -202              &      267\,475       &                            \tn
			                                                                                                                     & \nicefrac{5}{2} &        267\,394.7(2)         &      267\,395.7~\cite{Ryabtsev2006}       &      267\,993       &              -143               &             -203              &      267\,647       &                            \tn
			                                                        $8s$                                                         & \nicefrac{1}{2} &        268\,544.3(3)         & 268\,544\hphantom{.1}~\cite{Ryabtsev2006} &      269\,193       &              -132               &             -166              &      268\,895       &                            \tn
			                                                        $6f$                                                         & \nicefrac{7}{2} &        275\,919.8(3)         &                                           &      276\,430       &              -153               &             -201              &      276\,076       &                            \tn
			                                                                                                                     & \nicefrac{5}{2} &        276\,026.2(3)         &                                           &      276\,450       &              -152               &             -201              &      276\,097       &                            \tn
			                                                        $6g$                                                         & \nicefrac{9}{2} &        279\,863.6(2)         &                                           &      280\,580       &              -143               &             -202              &      280\,235       &                            \tn
			                                                                                                                     & \nicefrac{7}{2} &        279\,863.6(2)         &                                           &      280\,581       &              -143               &             -201              &      280\,237       &                            \tn
			                                                        $6h$                                                         &                 &        280\,067.8(7)         &                                           &                     &                                 &                               &                     &                            \tn
			                                                        $8d$                                                         & \nicefrac{3}{2} &        285\,265(1)\hphantom{.1}        &                                           &      285\,834       &              -140               &             -197              &      285\,497       &                            \tn
			                                                                                                                     & \nicefrac{5}{2} &        285\,370(1)\hphantom{.1}         &                                           &      285\,937       &              -143               &             -197              &      285\,597       &                            \tn
			                                                        $9s$                                                         & \nicefrac{1}{2} &                         & 286\,013\hphantom{.1}~\cite{Ryabtsev2006} &                     &                                 &                               &                     &                            \tn
			                                                        $7g$                                                         &                 &        292\,886.0(3)         &                                           &                     &                                 &                               &                     &                            \tn
			                                                        $7h$                                                         &                 &        293\,027.6(3)         &                                           &                     &                                 &                               &                     &                            \tn
			                                                        $7i$                                                         &                 &        293\,059.0(2)        &                                           &                     &                                 &                               &                     &                            \tn
			                                                 $10s$\hphantom{1}                                                   & \nicefrac{1}{2} &                         & 296\,844\hphantom{.1}~\cite{Ryabtsev2006} &                     &                                 &                               &                     &                            \tn
			                                                        $8g$                                                         &                 &        301\,338.2(6)         &                                           &                     &                                 &                               &                     &                            \tn
			                                                        $8h$                                                         &                 &        301\,439.2(7)         &                                           &                     &                                 &                               &                     &                            \tn
			                                                        $8i$                                                         &                 &        301\,462.3(7)         &                                           &                     &                                 &                               &                     &                            \tn \hline
			                                                         IP                                                          &                 & 328\,908.4(3) &  328\,550\hphantom{.1}~\cite{Moore1958}   &      329\,343       &              -143               &             -201              &      328\,999       &          327\,453          \tn
			                                                                                                                     &                 &                         & 328\,910\hphantom{.1}~\cite{Ryabtsev2006} &                     &                                 &                               &                     &
		\end{tabular}
	\end{ruledtabular}
\end{table*}

\subsection{$ l \leq l$\textsubscript{core} configurations}
The $5s$, $6s$, and $7s$ levels are included in Moore's tables~\cite{Moore1958}. The excitation energies of the $8s$, $9s$ and, $10s$ levels were determined in EUV spectroscopy experiments by Ryabtsev~{\it et al.}~\cite{Ryabtsev2006} in which transitions to the \mbox{$5p \; ^2\textrm{P}_{\textrm{\sfrac{1}{2}},\textrm{\sfrac{3}{2}}}$} terms were measured. The \mbox{$5p \; ^2\textrm{P}_{\textrm{\sfrac{1}{2}},\textrm{\sfrac{3}{2}}}$} terms can be populated from the $nd$ series, this $nd$ series is known up $n=7$. The highest known $np$ configuration so far was $6p$. All transition in the optical spectral range (Table \ref{tab:wavelength}) between both $ns$ and $nd$ and $6p$ agree with the literature excitation energies of the respective levels. 

The \mbox{$7p\; ^2\textrm{P}_{\textrm{\sfrac{1}{2}},\textrm{\sfrac{3}{2}}}$} levels are found by considering all possible transitions from the\mbox{ $7d\; ^2\textrm{D}_{\textrm{\sfrac{3}{2}},\textrm{\sfrac{5}{2}}}$} and \mbox{$8s\;^2\textrm{S}_{\textrm{\sfrac{1}{2}}}$} to the \mbox{$7p\; ^2\textrm{P}_{\textrm{\sfrac{1}{2}},\textrm{\sfrac{3}{2}}}$} levels. These levels can decay via emission in the visible to \mbox{$6d\; ^2\textrm{D}_{\textrm{\sfrac{3}{2}},\textrm{\sfrac{5}{2}}}$}. The excitation energies of the $7p$ configuration are established using seven transitions to three surrounding energy terms, providing a reliable assessment of the \mbox{$7p\; ^2\textrm{P}_{\textrm{\sfrac{1}{2}},\textrm{\sfrac{3}{2}}}$} level energies. For the \mbox{$7d\; ^2\textrm{D}_{\textrm{\sfrac{5}{2}}}$} good agreement with Ref.~\cite{Ryabtsev2006} is found, while for the \mbox{$7d\; ^2\textrm{D}_{\textrm{\sfrac{3}{2}}}$} level a difference of about 30\,\cm~is observed. The $ 8d $ is successively determined using two transitions to the $ 7p $, where the transition between equal angular momenta is likely too weak to be observed in our spectra. 

The quantum defects for $ l \leq l$\textsubscript{core} are shown in Fig.~\ref{quantumdefects}, calculated using the refined ionization limit. All data points exhibit a linear behavior, with only the lowest $5l$ configurations slightly deviating, reflecting a signature of the small quadratic term in Eq.~(\ref{eq:qd}). The most remarkable observation is the almost equal quantum defects of the \mbox{$5d\; ^2\textrm{D}_{\textrm{\sfrac{3}{2}},\textrm{\sfrac{5}{2}}}$} levels, indicative of an anomalously small fine structure interval of the \mbox{$5d\; ^2$D} term. This anomaly is further discussed in section~\ref{sec:fseffects}. 

\subsection{$ l > l$\textsubscript{core} configurations}
Of the high-$l$ configurations, i.e., $nf$, $ng$, $nh$, and $ni$ only the level energies of the \mbox{$4f\;^2$F} and \mbox{$5g\;^2$G} terms were known thus far. Our measurements confirm the small inverted fine structure splitting of approximately 60\,\cm~of the \mbox{$4f\;^2$F} term and of about 0.5\,\cm~of the \mbox{$5g\;^2$G} by direct comparison of the $ \Delta J=0 $ and $ \Delta J=-1 $ transitions. The \mbox{$4f\;^2$F} and \mbox{$5g\;^2$G} terms form the main basis onto which the excitation energies of the high-$l$ configurations are determined. Fine structure splittings of the \mbox{$ng\;^2$G} ($n \geq 6$), \mbox{$nh\;^2$H}, and \mbox{$ni\;^2$I} terms are too small to be determined, implying that the their fine structure splitting is less than 0.5\,\cm. The fine structure splittings of the $ nf $ terms are presented in Table~\ref{tab:fstabel} and will be discussed in detail in section~\ref{sec:fseffects}.

 \begin{table*}[t]
	\caption{Comparison of fine structure splittings in $ np $, $ nd $, and $ nf $ configurations of \mbox{Sn\,\textsc{iv}}. Experimental values stem from either a direct comparison of $\Delta J=0$ and $\Delta J=-1$ transitions or indirectly from the optimized level structure. The latter are labeled by an asterisk. The upper part of the Table contains results published in this work, the lower part shows results obtained from Refs.~\cite{Moore1958,Safronova2003,Ivanova2011,Ding2012,Grumer2014,Cheng1979}. \label{tab:fstabel}}
	\begin{ruledtabular}
		\begin{tabular}{lx{1.1cm}x{1.1cm}x{1.1cm}|x{1.1cm}x{1.1cm}x{1.1cm}x{1.1cm}|x{1.1cm}x{1.1cm}x{1.1cm}}
			 & \multicolumn{10}{c}{fine structure splitting (\cm)}                  \tn \cline{2-11}
			 & $ 5p $ & $ 6p $ & $ 7p $ & $ 5d $ & $ 6d $ & $ 7d $ & $ 8d $ & $4f$ & $5f$ & $ 6f $  \tn \hline
			experiment  &  & 2\,179.5 & \hphantom{\,1}909.1 & \hphantom{.0}107.0 & \hphantom{.1}331.1 & 179.3 & $105.4^{*}$ & \hphantom{1}-60.4 & -$309.6^{*}$ & -$106.4^{*}$    \tn
			\cowan &6\,417\hphantom{.4}  & 2\,237\hphantom{.1} & \hphantom{1\,}911\hphantom{.1} & 170 &240 &  130\hphantom{.3}& \hphantom{1}79\phantom{.5*} &34  &-228\phantom{.5*}  &\hphantom{1}-73\phantom{$.5^*$}  \tn
			FSCC & 6\,515\hphantom{.1} & 2\,191\hphantom{.1} & 1\,139\hphantom{.1} & 736 & 326 & 172\hphantom{.3} & 100\phantom{.5*} & \hphantom{1}-73\hphantom{.4} & \hphantom{1}-40\phantom{.5*} & \hphantom{1}-21\phantom{$.5^*$}     \tn
			CI+MBPT &  &  &  & 162 &  &  &  &  & -620\phantom{.5*} &    \tn \hline
			experiment~\cite{Moore1958} & 6\,508.4 & 2\,179.9 &  & \hphantom{.0}106.1 & \hphantom{.1}332.0 &  &  & \hphantom{1}-60.5 &  &     \tn 
			RMBPT~\cite{Safronova2003} & 6\,471\hphantom{.1} &  &  & 745 &  &  &  & \hphantom{1}-76\hphantom{.4} & \hphantom{1}-44\phantom{.5*} &     \tn
			RPTMP\footnote{Fine structure splittings deduced from transition wavelengths. \label{fscalc}}~\cite{Ivanova2011} &  &  &  &  &  &  &  & \hphantom{1}-60\hphantom{.4} & \hphantom{1}-22\phantom{.5*} &      \tn
			MCDHF~\cite{Ding2012} &  &  &  &  &  &  &  & \hphantom{1}-71\hphantom{.4} &  &     \tn
			FCV~\cite{Grumer2014} &  &  &  &  & &  &  & \hphantom{1}-85\hphantom{.4} &  &     \tn
			RHF\textsuperscript{\ref{fscalc}}~\cite{Cheng1979} & 5\,960\hphantom{.1} &  &  & 641 &  &  &  & -108\phantom{.5} &\hphantom{1}-72\phantom{.5*} & 
		\end{tabular}
	\end{ruledtabular}
\end{table*}

The first level of the $nh$ series, $6h$, is found by assigning the strong transition from this level to $5g$. The $6h$ is the lower level of the transitions determining the $7i$ and $8i$. The $7h$, $8h$ are found by transitions to the $6g$, which is based on the transition to the \mbox{$5f\;^2$F}.The \mbox{$ng\;^2$G} ($n \geq 7$) are determined from their transitions to the \mbox{$5f \;^2$F} and \mbox{$6f \;^2$F} terms.The $ 5f $ and $ 6f $ terms are defined by transitions to their lower-lying \mbox{$nd \;^2$D} counterparts. 

The relative values of the quantum defects for the $ng$, $nh$, and $ni$ series are in good agreement with the $nl$ scaling laws for $ l > l$\textsubscript{core} as presented by Edl\'en~\cite{Edlen1964}. The quantum defects for the $nf$ series are about a factor of three to four larger than expected from these scaling laws.
In addition, relatively large fine structure splittings are observed for the $5f$ and \mbox{$6f \;^2$F} terms.
Both effects may be a signature of an enhanced interaction with core-electron configurations. 

\subsection{Anomalous fine-structure effects in the $5d\;^2\textnormal{D}$ and $nf\;^2\textnormal{F}$ terms \label{sec:fseffects}}
Table \ref{tab:fstabel} summarizes experimental and theoretical fine structure intervals in \mbox{Sn\,\textsc{iv}}. We have performed Fock space coupled cluster and configuration interaction many-body perturbation theory in order to address the aforementioned anomalous values of the fine-structure intervals in the \mbox{$5d\;^2$D} and \mbox{$nf \;^2$F} terms.

The FSCC calculations of the transition energies were performed within the framework of the projected Dirac-Coulomb-Breit (DCB) Hamiltonian~\cite{Suc80},
\begin{eqnarray}
H_{DCB}= \displaystyle\sum\limits_{i}h_{D}(i)+\displaystyle\sum\limits_{i<j}(1/r_{ij}+B_{ij}).
\label{eqHdcb}
\end{eqnarray}
Here, $h_D$ is the one electron Dirac Hamiltonian,
\begin{eqnarray}
h_{D}(i)=c\boldsymbol{\alpha}_{i}\cdot \mathbf{p}_{i}+c^{2}\beta _{i}+V_{nuc}(i),
\label{eqHd}
\end{eqnarray}
where $\bm{\alpha}$ and $\beta$ are the four-dimensional Dirac matrices.  The nuclear potential $V_{nuc}(i)$  takes into account the finite size of the nucleus, modeled by a uniformly charged sphere~\cite{IshBarBin85}. The two-electron term includes the non-relativistic electron repulsion and the frequency independent Breit operator,
\begin{eqnarray}
B_{ij}=-\frac{1}{2r_{ij}}[\boldsymbol{\alpha }_{i}\cdot \boldsymbol{\alpha }_{j}+(%
\boldsymbol{\alpha }_{i}\cdot \mathbf{r}_{ij})(\boldsymbol{\alpha }_{j}\cdot \mathbf{%
	r}_{ij})/r_{ij}^{2}],
\label{eqBij}
\end{eqnarray}
and is correct  to second order in the fine structure constant $\alpha$.

The calculations of the transition energies of Sn\textsuperscript{3+} start from the closed-shell reference \mbox{[Kr]$4d^{10}$} configuration of Sn\textsuperscript{4+}. After the first stage of the calculation, consisting of solving the relativistic Hartree-Fock equations and correlating the closed-shell reference state, a single electron was added to reach the desired Sn\textsuperscript{3+} state. A large model space was used in this calculation, comprising 10\,$s$, 8\,$p$, 6\,$d$, 6\,$f$, 4\,$g$, 3\,$h$, and 2\,$i$ orbitals in order to obtain a large number of excitation energies and to reach optimal accuracy. The intermediate Hamiltonian method was employed to facilitate convergence~\cite{EliVilIsh05}. 

The uncontracted universal basis set~\cite{MalSilIsh93} was used, consisting of even-tempered Gaussian type orbitals, with exponents given by
\begin{eqnarray}
\xi _{n} &=&\gamma \delta ^{(n-1)},\text{ \ \ }\gamma =106\,111\,395.371\,615, \\
\delta  &=&0.486\,752\,256\,286. \nonumber
\label{eqUniversal}
\end{eqnarray}

The basis set was composed of 37\,$s$, 31\,$p$, 26\,$ d $, 21\,$ f $, 16\,$ g $, 11\,$ h $, and 6\,$ i $ functions; the convergence of the obtained transition energies with respect to the size of the basis set was verified. All the electrons were correlated.

The FSCC calculations were performed using the Tel-Aviv Relativistic Atomic FSCC~code~(\textsc{trafs-3c})~\cite{trafs-3c}. To account for the QED corrections to the transition energies we applied the model Lamb shift operator (MLSO) of Shabaev and co-workers~\cite{ShaTupYer15} to the atomic no-virtual-pair many-body DCB Hamiltonian as implemented into the \textsc{qedmod} program. Our implementation of the MLSO formalism into the Tel Aviv atomic computational package allows us to obtain the vacuum polarization and self-energy contributions beyond the usual mean-field level, namely at the DCB-FSCCSD level.

The FSCC results are compared to the experimental level energies and several results of previous theoretical work in Table~\ref{tab:energylevels} and are overall in good agreement. Typical differences with experiment are about 100 to 300~\cm~which is on the 10\textsuperscript{-3} level of the calculated excitation energies. Concerning the measured anomalous fine structure intervals of the \mbox{$5d\;^2$D} and \mbox{$5f, 6f \;^2$F} terms, presented in Table~\ref{tab:fstabel}, the apparent narrowing of the fine-structure interval of the $5d\;^2$D term and the widening of the \mbox{$5f, 6f \;^2$F} term intervals are not reproduced by the FSCC calculations. The FSCC intervals are similar to those presented in earlier theoretical investigations~\cite{Safronova2003,Ivanova2011,Ding2012,Grumer2014,Cheng1979}.

For the \mbox{$5d\;^2$D} term the fine structure interval is measured at 107\,\cm, while all theoretical results are higher by a factor of approximately seven, see Table~\ref{tab:fstabel}. When inspecting the level diagram (Fig.~\ref{fig:leveldiagram}) one notices that the \mbox{$5d\;^2$D} term might suffer from configuration interaction of the doubly-excited \mbox{$4d\textsuperscript{9}\,5s^2$} levels.

To quantify the strength of the configuration interaction we employ configuration interaction many-body perturbation theory (CI+MBPT) calculations using the \ambit~code. Details of the \ambit~code can be found in Refs.~\cite{Berengut2006,Berengut2011,Berengut2016,Kahl2018}. To begin our discussion of the \ambit~treatment of the problem, first consider the \mbox{$5d\;^2\textrm{D}_{\textrm{\sfrac{5}{2}}}$} and \mbox{$4d\textsuperscript{9}\,5s^2\; ^2\textrm{D}_{\textrm{\sfrac{5}{2}}}$} levels as a two-level system. In the absence of interaction between them, they have theoretical energies $\epsilon_1$ and $\epsilon_2$, respectively. $\epsilon_1$ is, to a good approximation, the FSCC value of the \mbox{$5d\;^2\textrm{D}_{\textrm{\sfrac{5}{2}}}$} level, since in that calculation the one-hole two-particle \mbox{$4d\textsuperscript{9}\,5s^2\; ^2\textrm{D}_{\textrm{\sfrac{5}{2}}}$} level is not explicitly included. If we now add an interaction $V$, then the states mix and the levels repel each other. 

The Hamiltonian of this two-level system is
\[
H = \left( \begin{array}{cc} \epsilon_1 & V \\ V & \epsilon_2 \end{array} \right) \ .
\]
Writing $\Delta\epsilon = \epsilon_2 - \epsilon_1 > 0$, the \mbox{$5d\;^2$D$_{\textrm{\sfrac{5}{2}}}$} level shifts down by an amount
\begin{equation}
\label{eq:E_shift_b}
\delta =
\frac{\Delta E - \Delta\epsilon}{2} =
\frac{\Delta\epsilon}{2} \left( \sqrt{1 + \frac{4V^2}{\Delta\epsilon^2}} - 1 \right)
=b^2 \Delta E,
\end{equation}
where $\Delta E = E_2 - E_1$ is the difference between the eigenvalues of $H$ and $b$ is the smaller component of the normalized eigenvector $(a, b)^\intercal$. 

Using \ambit, we calculate theoretical values for the parameters $\Delta E_\textrm{th}$ and $b$, from which we can obtain the interaction \mbox{$V = -ab\Delta E_\textrm{th}$}. However, the values of $\Delta E_\textrm{th}$ and $b$ are sensitive to details of the calculations. In particular, it is challenging to match theoretical with experimental level energies at a good level of accuracy. On the other hand, the values of $V$ that we obtain are highly stable since they are not sensitive to the separation. If we use the experimental separation \mbox{$\Delta E_\textrm{exp} = 3\,823\,\cm$} and \mbox{$|V| = 1\,523\,\cm$} from \ambit, we find the energy shift of the \mbox{$5d\;^2$D$_{\textrm{\sfrac{5}{2}}}$} level due to interaction with the hole state,
\begin{equation}
\label{eq:E_shift}
\delta = \frac{\Delta E_\textrm{exp}}{2} \left( 1 - \sqrt{1 - \frac{4V^2}{\Delta E_\textrm{exp}^2}} \right) ,
\end{equation}
yielding \mbox{$-755\,\cm$}. The \mbox{$5d\;^2\textrm{D}_{\textrm{\sfrac{3}{2}}}$} level also shifts down due to interaction with the \mbox{$4d^{9}\,5s^2\ ^2\textrm{D}_{\textrm{\sfrac{3}{2}}}$} hole level. However, the energy difference is three times larger, and since $b \sim 1/\Delta E$, the energy shift is smaller by approximately a factor of three. We calculate $|V| = 1\,498\,\cm$ for this pair of levels, so Eq.~(\ref{eq:E_shift}) gives a level shift of $-181\,\cm$. The change in the $5d$ fine-structure splitting is therefore $-574\,\cm$ which is close to the difference between experiment and FSCC calculation of $-629\,\cm$.

In support of the role of configuration interaction on the \mbox{$5d\;^2\textrm{D}$} fine structure, FSCC calculations were performed for isoelectronic \mbox{In$^{2+}$} ions in which there is a much larger energy difference between the \mbox{$5d\;^2$D} term and the doubly-excited \mbox{$4d\textsuperscript{9}\,5s^2 \; ^2\textrm{D}_{\textrm{\sfrac{5}{2}}}$} level~\cite{Moore1958}. Thus a better agreement with FSCC calculations is expected. In comparison to Sn$^{3+}$, for In$^{2+}$ the difference between experiment~\cite{Moore1958} and FSCC indeed reduces strongly from a factor of seven to only 30\% (298 versus 398\,\cm).

Before discussing the impact of configuration interaction on the fine structure of the \mbox{$5f\;^2$F} and \mbox{$6f\;^2$F} terms we note that the \mbox{$4f\;^2$F} exhibits an inverted fine structure with the $J=\nicefrac{7}{2}$ level being more stronger bound than the $J=\nicefrac{5}{2}$ level by approximately 60\,\cm. The occurrence of this inversion of the fine structure and the actual value of the fine-structure interval results from an intricate balance between relativistic, spin-orbit and core polarization effects and has been the subject of a variety of theoretical approaches calculating the \mbox{$4f\;^2$F} fine-structure along the isoelectronic sequence of Ag-like ions~\cite{Safronova2003,Ivanova2011,Ding2012,Grumer2014,Cheng1979}. 

We confirm the fine structure of the \mbox{$4f\;^2$F} term by measuring the wavelengths of the transitions from the \mbox{$5g\;^2$G} levels and the transitions to the \mbox{$5d\,^2$D} levels. Weaker transitions from the \mbox{$6d\;^2$D} term to the doubly-excited \mbox{$4d^9\,5s^2\;^2\textrm{D}_{\textrm{\sfrac{5}{2}}}$} level are observed additionally. A comparison of the fine structure splitting of the observed \mbox{$nf\;^2$F} levels with theoretical calculations is given in Table~\ref{tab:fstabel}. Both measurements and theoretical calculations agree on an inverted fine structure splitting for these \mbox{$nf\;^2$F} terms. However the magnitude of the fine structure splitting of the $ 5f $ and $ 6f $ terms are much smaller than our experimental ones. 

In similar fashion to the \mbox{$5d\;^2$D} levels, the \mbox{$5f\;^2$F} fine-structure splitting is strongly affected by interaction with hole states. However, the \mbox{$5f\;^2$F} case is more complicated because the \mbox{$4d\textsuperscript{9}\,5s\,5p$} configuration has seven configuration state functions (CSFs) with $J=\nicefrac{5}{2}$ and four CSFs with $J=\nicefrac{7}{2}$. The CSFs tend to be strongly mixed with each other, and also have small contributions of CSFs belonging to other configurations. Therefore, rather than treat the system as a few-level system, we use the approach of perturbation theory.

Using \ambit~we obtain energies for the \mbox{$4d\textsuperscript{9}\,5s\,5p$} levels as well as a mixing coefficient. At first order in perturbation theory, the coefficient of $\psi_{5f}$ in the level $\psi_i$ is simply
\[
b_{5f} = \frac{V_{i, 5f}}{E_i - E_{5f}} .
\]
From our \ambit~values of $b_{5f}$ and $E_i$, we extract values for the matrix element $V_{i, 5f}$. Again these are relatively stable for different calculations, even though the energies and $b$ coefficients can change dramatically.

The corresponding energy shift of the $5f$ level is $\delta_{5f} = V_{i, 5f}^2/\Delta E_\textrm{exp}$. Unfortunately we do not have precise experimental determinations of most of the interaction \mbox{$4d\textsuperscript{9}\,5s\,5p$} levels. Instead we use the results of \cowan~calculations (Ref.~\cite{Ryabtsev2006}) to obtain an approximation to the level shifts. The results are presented in Table~\ref{tab:5f}. 

\begin{table}[htb]
	\caption{\mbox{Sn\,\textsc{iv}} \mbox{$4d^{9}\, 5s\, 5p$} level energies as candidates for possible configuration interaction with the \mbox{$ nf$} levels. The level energies are obtained from \cowan~code calculations published in Ref.~\cite{Ryabtsev2006}. The matrix elements $|V_{i,5f}|$ are calculated using the \ambit~code. The resulting shift of the $5f$ level by configuration interaction by the level is given by $\delta_{5f}$. The $J=\nicefrac{5}{2}$ levels interacts with the \mbox{$5f\;^2\textrm{F}_{\textrm{\sfrac{5}{2}}}$}, similarly the $J=\nicefrac{7}{2}$ levels with \mbox{$5f\;^2\textrm{F}_{\textrm{\sfrac{7}{2}}}$}.  \label{tab:5f}}
	\begin{ruledtabular}
		\begin{tabular}{llrrr}
			$E_i$\,(\cm)~\cite{Ryabtsev2006}                                                   & $J$ & $|V_{i,5f}|$ & $\delta_{5f}$ &  \\ \hline
			226\,363                                                                                         & 5/2 &            3 &             0 &  \\
			231\,318                                                                                         & 5/2 &          484 &            11 &  \\
			239\,582                                                                                         & 5/2 &          765 &            47 &  \\
			242\,203                                                                                         & 5/2 &          622 &            39 &  \\
			249\,541                                                                                         & 5/2 &           41 &             1 &  \\
			263\,718                                                                                         & 5/2 &          111 &            -1 &  \\
			269\,440                                                                                         & 5/2 &         1341 &          -104 &  \\ \hline
			231\,090                                                                                         & 7/2 &           72 &             0 &  \\
			239\,920                                                                                         & 7/2 &          417 &            15 &  \\
			246\,851                                                                                         & 7/2 &          601 &            72 &  \\
			260\,398 & 7/2 &         2404 &          -676 &
		\end{tabular}
	\end{ruledtabular}
\end{table}

We see that while each of the $5f$ levels are shifted by interactions with the hole levels, the change in the fine-structure splitting is dominated by the interaction of the \mbox{$5f\; ^2\textrm{F}_{\textrm{\sfrac{7}{2}}}$} level with a hole state at 260\,398\,\cm. The final expected shift is 580\,\cm, overestimating the actual difference between experimental fine-structure and FSCC calculations of 270\,\cm. Nevertheless, given the uncertainties in our estimation of $V$ and the location of the doubly-excited levels, we arrive at a plausible explanation for the observed anomaly.

A similar explanation can be given for the observed difference in fine structure splitting for the \mbox{$6f\; ^2\textrm{F}$} term, interacting with the high-lying levels in the same series of hole states. Because the energy differences between these levels are larger, this effect may reasonably be expected to be smaller than for the \mbox{$5f\; ^2\textrm{F}$} levels. Likewise, the \mbox{$7p\; ^2\textrm{P}$} is expected to interact with several doubly-excited levels. 

We also performed \cowan~calculations to investigate the terms described above. The number of fitted parameters in this case is reduced by tying the Hartree-Fock ratios to the spin-orbit parameters for the $np$, $nd$, and $nf$ levels. The \mbox{$4d^{9}\,5s\,5p$} levels are taken from Ref.~\cite{Ryabtsev2006} (included in Table~\ref{tab:5f}). The two \mbox{$4d^{9}\,5s^2 \; ^2\textrm{D}$} levels were calculated with two adjustable parameters. All interaction parameters were fixed at 0.8 of their Hartree-Fock values. The number of levels determined was insufficient to more accurately fit the parameters. Results of the calculation are included in Table~\ref{tab:fstabel} and show good agreement with the experimental results, although the inversion of the \mbox{$4f\; ^2\textrm{F}$} level is not reproduced. The agreement for all other $nf$ and the $np$ and $nd$ levels underlines the significant role of configuration interaction in a quasi-one-electron system like Sn$^{3+}$.

\section{Conclusion}
Optical techniques are useful diagnostics in plasma sources of EUV light in nanolithography. We present the ultraviolet and optical spectra of a laser-produced tin plasma. The lines belonging to Sn\textsuperscript{3+} are identified using a convenient masking technique. The 33 newly found lines are used to determine 13 new configurations with iterative guidance from \cowan~code calculations. The level energies are verified using quantum-defect scaling procedure, leading to the refinement of the ionization limit to 328\,908.4\,\cm~with an uncertainty of 2.1\,\cm. FSCC calculations are generally in good agreement with present measurements. The anomalous behavior of the \mbox{$ 5d\; ^2\textrm{D} $} and \mbox{$ nf\;^2\textrm{D}$} terms is shown to arise from configuration interaction with doubly-excited levels by joining the strengths of FSCC, \cowan~and CI+MBPT approaches.

\begin{acknowledgments}	
Part of this work has been carried out at the Advanced Research Center for Nanolithography, a public-private partnership between the University of Amsterdam, the Vrije Universiteit Amsterdam, the Netherlands Organization for Scientific Research (NWO), and the semiconductor equipment manufacturer ASML. 
\end{acknowledgments}


%

\end{document}